\documentclass[slac_one]{revtex4}
\usepackage{graphicx}
\usepackage{fancyhdr}
\begin{document}

\title{Reconstruction of Cosmic and Beam-Halo Muons with the CMS Detector}

%

\author{N. Neumeister}
\affiliation{Purdue University, W. Lafayette, IN 47907, USA}
\author{C. Liu}
\affiliation{Purdue University, W. Lafayette, IN 47907, USA}

\begin{abstract}
The powerful muon and tracker systems of the CMS detector together with dedicated reconstruction software allow precise and efficient measurement of muon tracks originating from proton-proton collisions. The standard muon reconstruction algorithms, however, are inadequate to deal with muons that do not originate from collisions. We present the design, implementation, and performance of a dedicated cosmic muon track reconstruction algorithm, which features pattern recognition optimized for muons that are not coming from the interaction point, i.e. cosmic muons and beam-halo muons. To evaluate the performance of the new algorithm, data taken during Cosmic Challenge phases I and II as well as beam-halo muons recorded during the first LHC beam operation were studied. In addition, a variety of more general topologies of cosmic muons and beam-halo muons were studied using simulated data to demonstrate some key features of the new algorithm.
\end{abstract}

\maketitle

\thispagestyle{fancy}

\section{INTRODUCTION} 
The efficient and accurate detection of muons and the reconstruction of their momenta with
high precision over a large range of muon energies are crucial for the LHC physics program. The
Compact Muon Solenoid (CMS) experiment~\cite{cmsproposal} at the LHC provides excellent muon
identification and reconstruction capabilities. A large superconducting solenoid with a 4~T
magnetic field provides strong bending power, allowing a precise measurement of the momentum. A
complex muon system has been designed that consists of 3 different types of detectors,
sandwiched between layers of the iron return yoke. Centrally-produced muons are detected in the
silicon tracker, the calorimeters, and the muon system.

In addition to centrally-produced muons, particles that do not
originate from $p$-$p$ collisions, such as cosmic muons and beam-halo muons, can be recorded by
the CMS detector. However, the detection and reconstruction of cosmic and beam-halo muons are
different from that of muons from $p$-$p$ collisions. Cosmic muons are the most abundant
particles originating from cosmic rays at sea level~\cite{pdg}. Beam-halo muons are machine-induced
particles that travel along the beam line. Although in physics analyses these types of muons are generally
considered as sources of background, they can be used for detector
alignment, calibration, and detector performance validation. The efficient reconstruction of
cosmic and beam-halo muons is especially important for the commissioning phase of the detector.

\begin{figure}
\begin{center}
$\begin{array}{c@{\hspace{0.1in}}c}
\resizebox{0.4\hsize}{!}{\includegraphics*{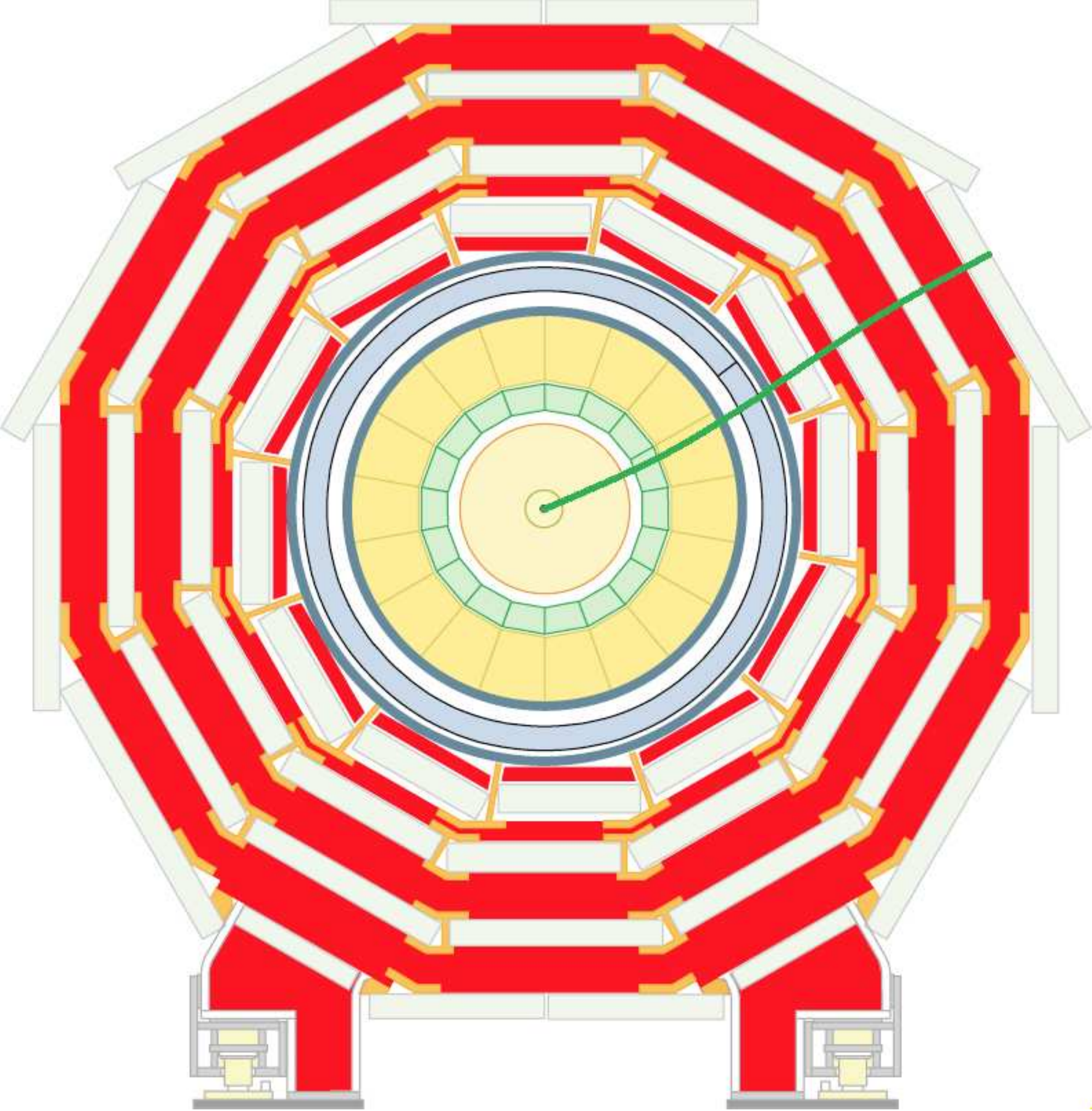}} &
\resizebox{0.4\hsize}{!}{\includegraphics*{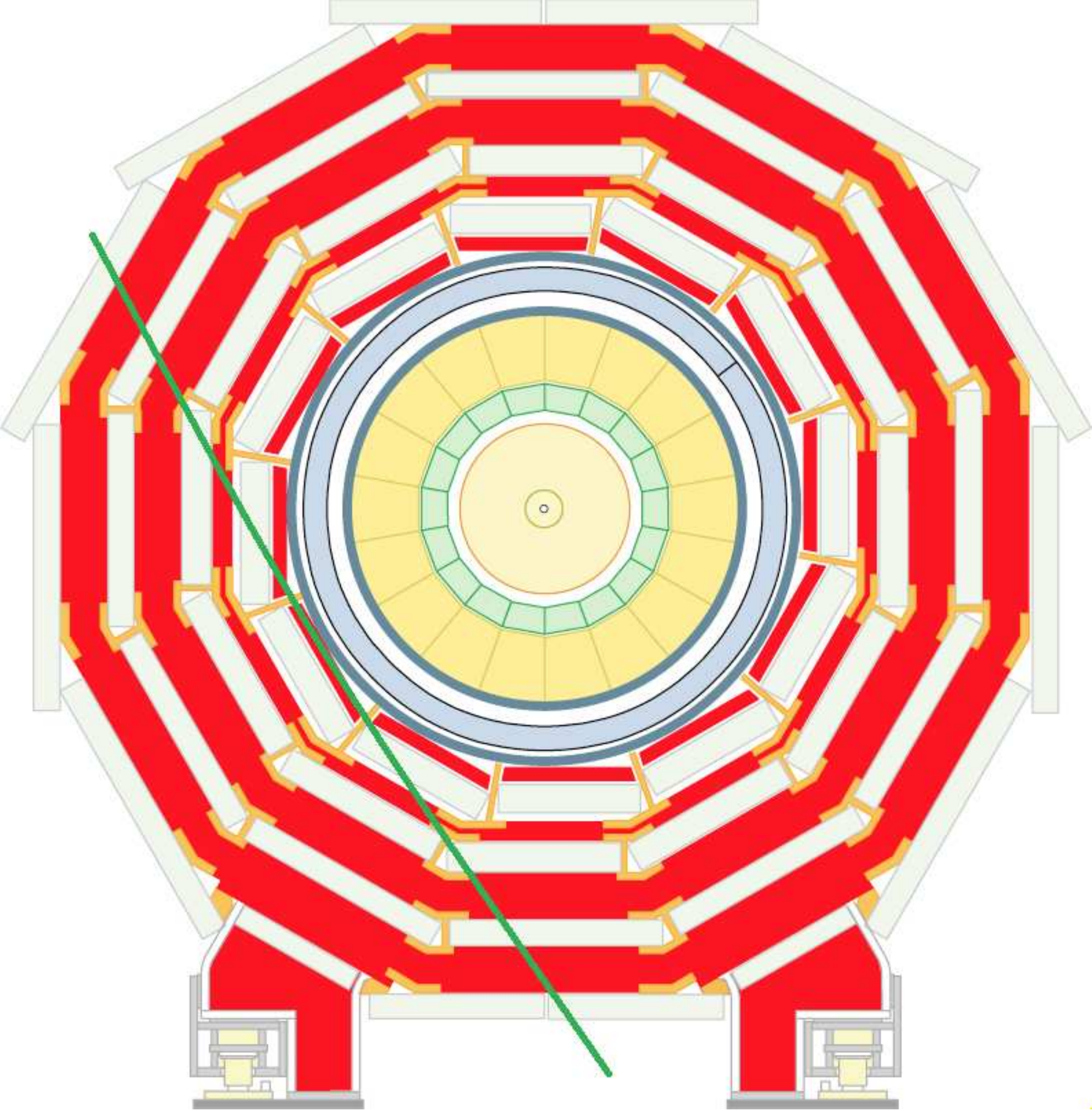}} \\ [0.1cm]
\mbox{\bf (a)} & \mbox{\bf (b)} \\
\resizebox{0.4\hsize}{!}{\includegraphics*{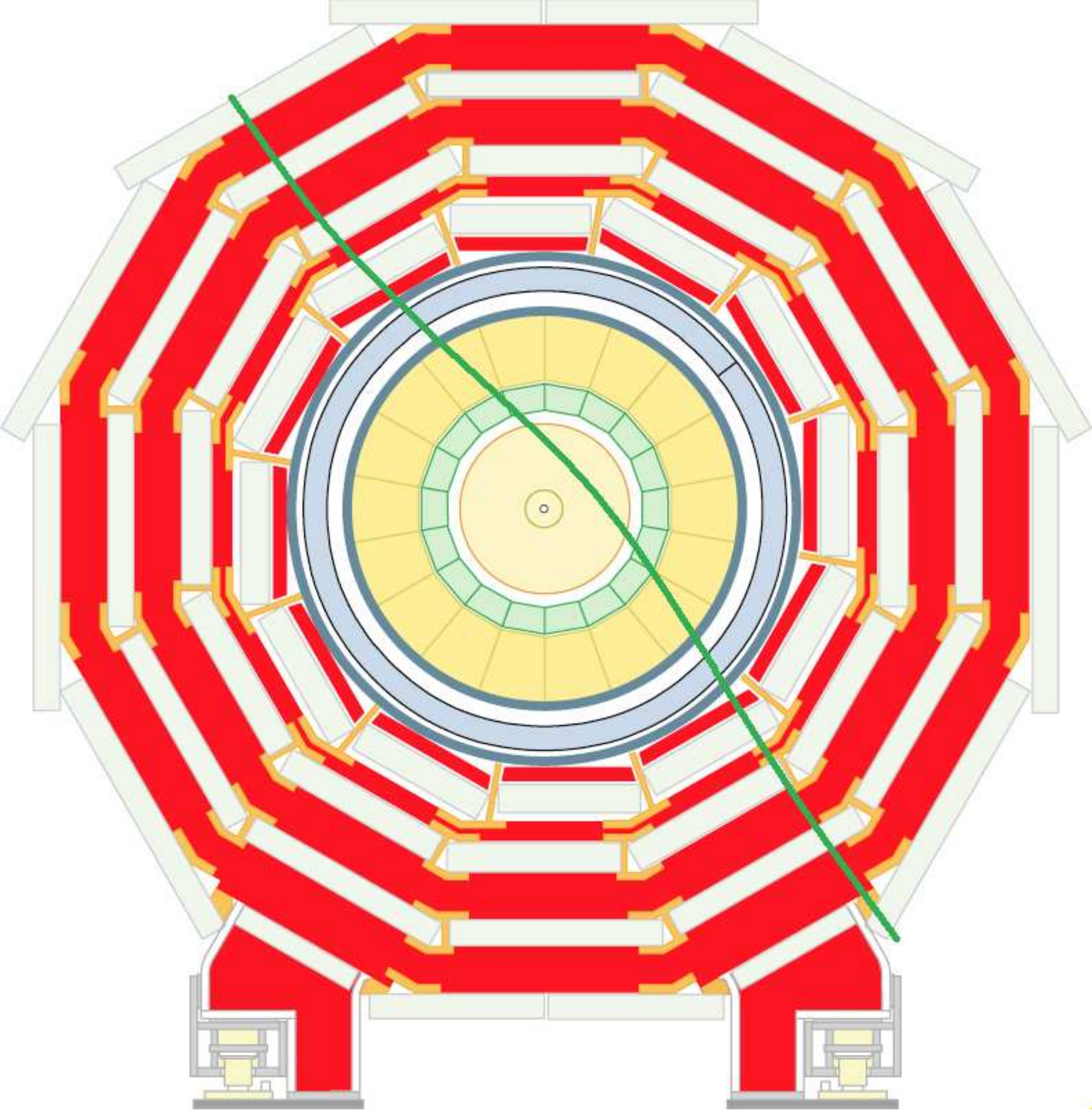}} &
\resizebox{0.4\hsize}{!}{\includegraphics*{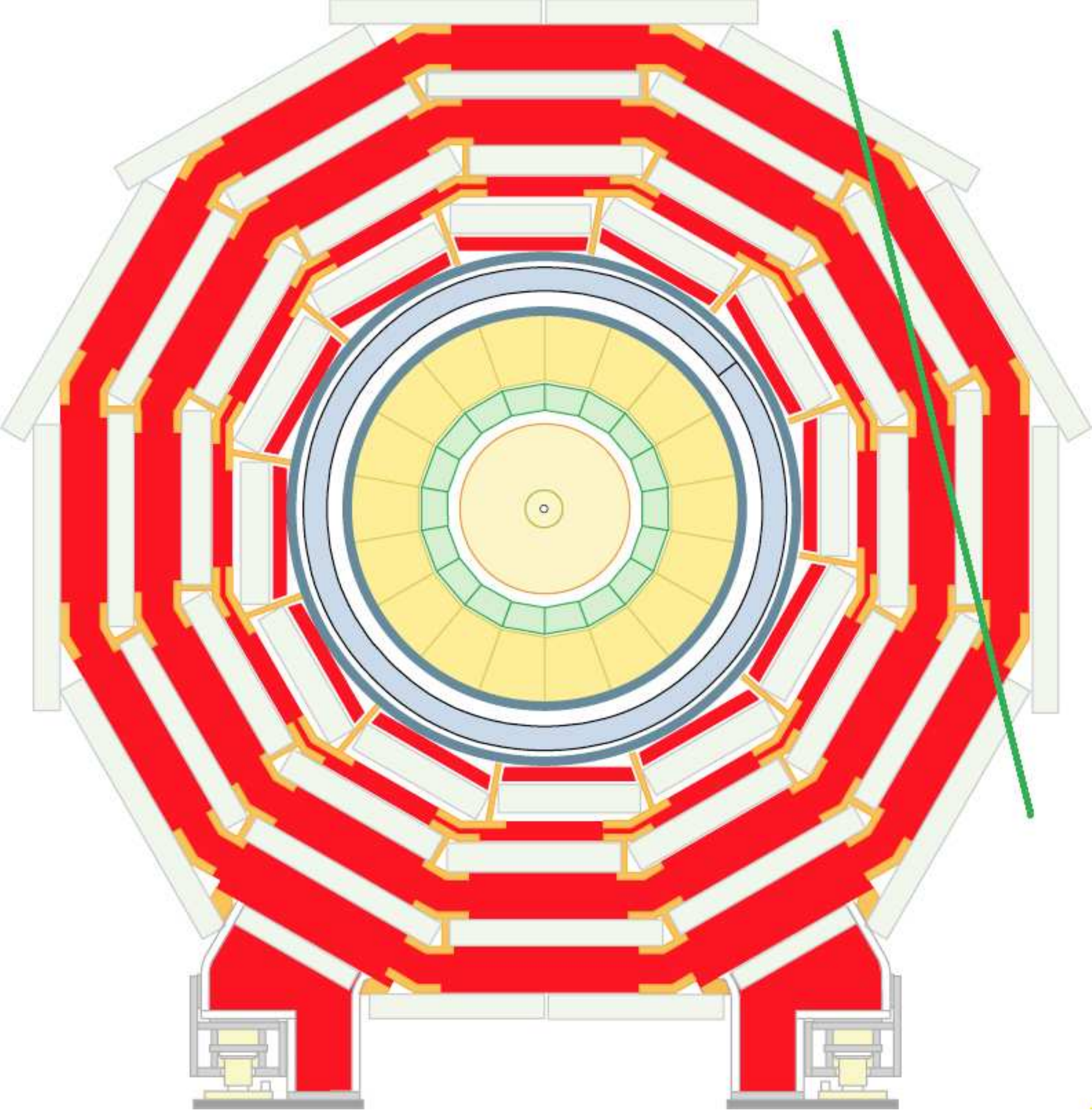}} \\ [0.1cm]
\mbox{\bf (c)} & \mbox{\bf (d)} \\
\resizebox{0.5\hsize}{!}{\includegraphics*{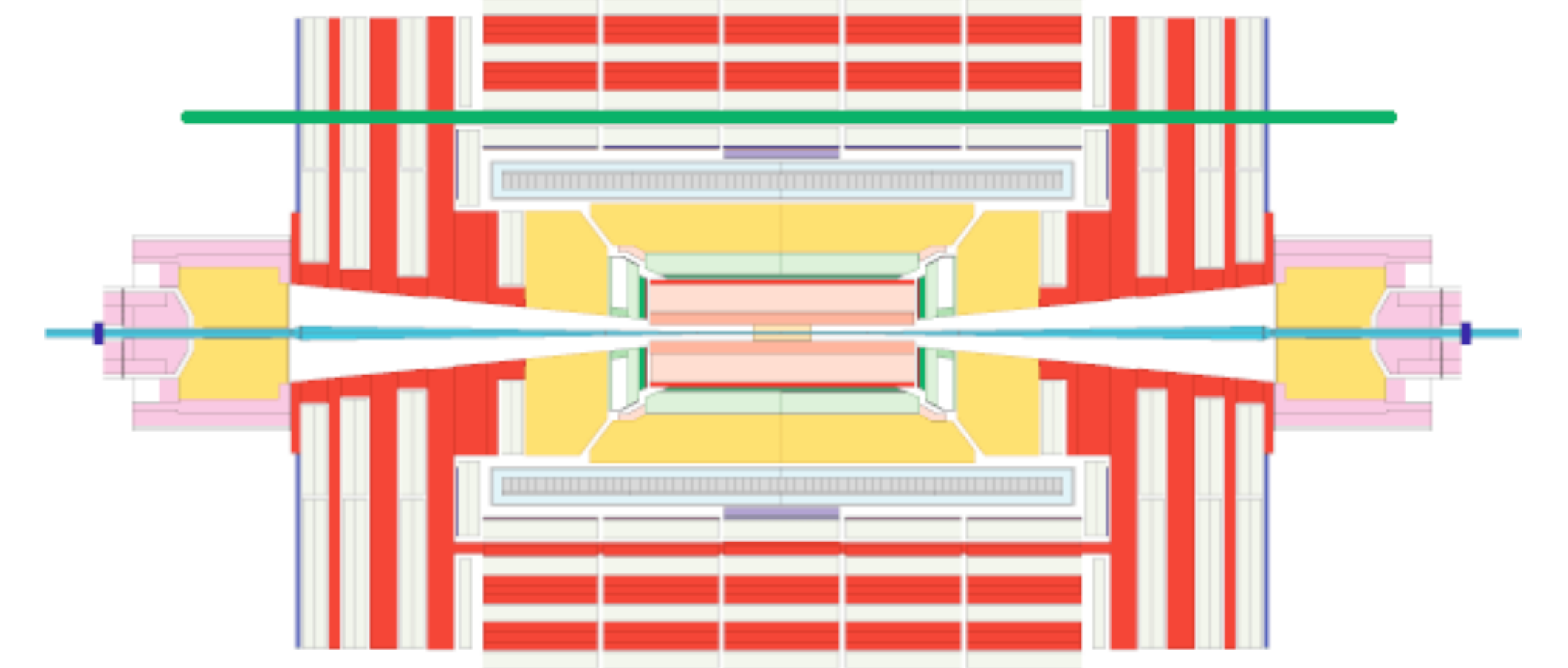}} &
\resizebox{0.5\hsize}{!}{\includegraphics*{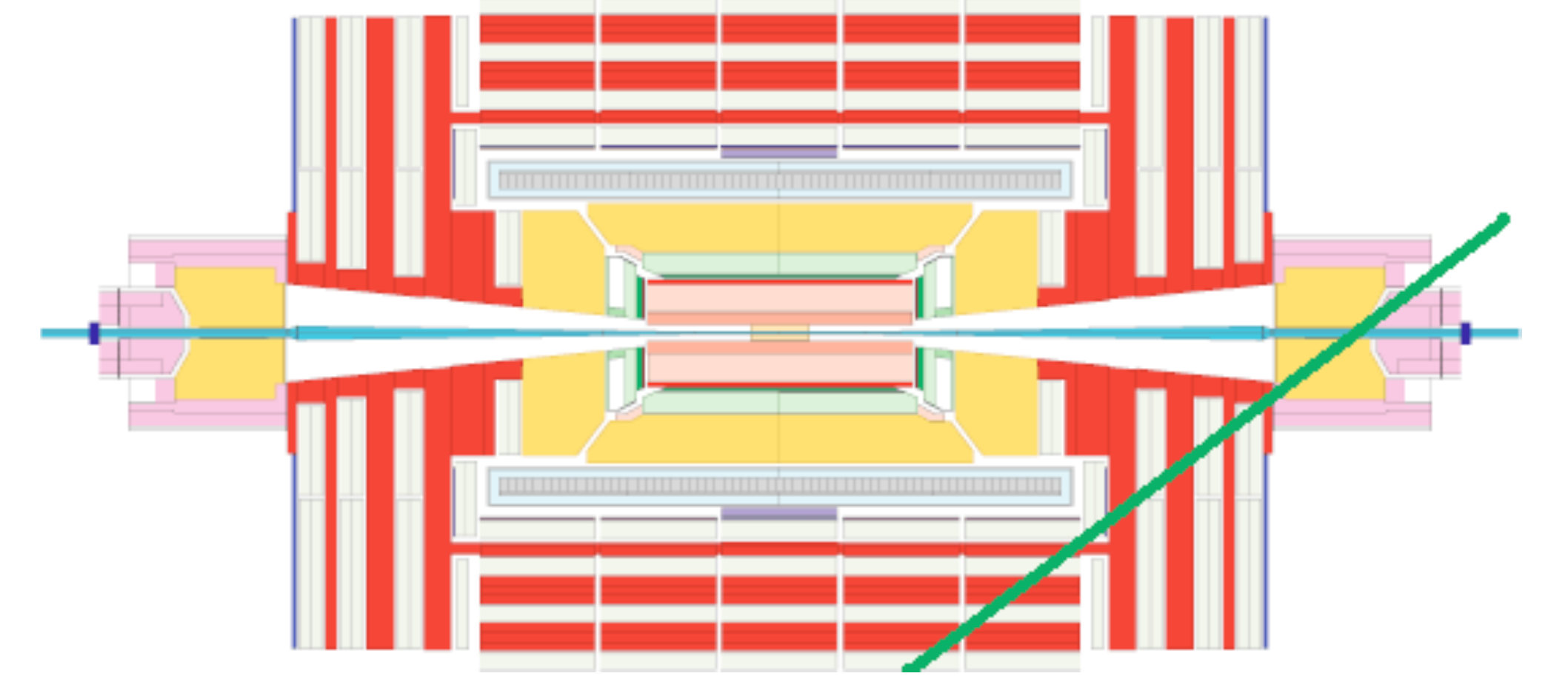}} \\ [0.1cm]
\mbox{\bf (e)} & \mbox{\bf (f)} \\
\end{array}$
\end{center}
\caption{Illustration of the differences among muons from $p$-$p$ collisions,
different types of cosmic muons, and beam-halo muons. 
(a) Muons from collisions always propagate from the center to the
outside and the pattern is well-defined; (b) Cosmic muons can penetrate the
detector and leave signals in opposite hemispheres of the muon system; 
(c) Cosmic muons can leave signals in the tracker system and
 opposite hemispheres of the muon system;
(d) Cosmic muons can enter the 
detector and leave without passing through all muon detector layers; (e) beam-halo
muons can penetrate the detector and leave signals
in both endcap regions;
(f) Cosmic muons can enter the endcap region and leave from the barrel region
of the detector (or vice versa, in the upper part of the detector).}
\label{cospp}
\end{figure}

Since the standard muon reconstruction software has been optimized to identify and reconstruct
muons originating from $p$-$p$ collisions, a different optimization must be carried out 
to reconstruct effectively the muons coming from outside the detector. A dedicated cosmic muon
reconstruction software was developed and the performance was tested with real data taken during the CMS Magnet
Test and Cosmic Challenge (MTCC)~\cite{mtcc}. Unlike muons from collisions, which are moving radially
outward, cosmic muons arrive at the detector from random directions and at random times. They
can traverse either both hemispheres or only a small part of the detector depending on their
energy and direction. Figure~\ref{cospp} illustrates the different topologies of muons coming
from outside and from $p$-$p$ collisions. In some cases, as indicated in
Figs.~\ref{cospp}(b) and (c), the standard muon reconstruction algorithm can reconstruct a
cosmic muon, but the muon will be recognized as 2 separate tracks. Cosmic muons arriving at the detector in coincidence with $p$-$p$ collisions are
a potential background for the physics processes. Distinguishing them from real muon events is
crucial for many physics analyses. In addition, reconstructing such muon trajectories
provides an important tool for aligning detector components and studying trigger and
reconstruction efficiencies, especially during the initial data taking
period~\cite{drollinger}.

Muon reconstruction as implemented in the official CMS software framework
is performed in 3 stages:
local pattern recognition within each muon chamber, standalone reconstruction that builds
tracks within the muon system, and global reconstruction that builds tracks using data from
the muon system and the silicon tracker. Already at the level of local reconstruction in the
muon system, cosmic muons and beam-halo muons should be treated differently. For example, in
the barrel drift tube muon system, drift times are recorded and transformed to local positions for further
reconstruction. The latencies of different drift tube chambers and readout electronics are
different and depend on the location of the muon track and on its time of arrival within the (arbitrary, in the case of cosmic muon) bunch crossing (BX) window
defined by the trigger. Since cosmic muons arrive randomly in time, a specific calibration
process is carried out as discussed in~\cite{philipp,dtcalib}.  In this article, we focus on the standalone and
global muon reconstruction steps by presenting the limitation of the standard reconstruction
algorithms and proposing an alternative reconstruction algorithm for cosmic and 
beam-halo muons. The standard algorithms are designed with the assumption that muons are coming
from the interaction point and the direction of the energy flow of trajectories is always
out-going from the center of the detector. Pattern recognition based on this assumption is
not suitable for the reconstruction of muons coming from outside the detector, except for some
special cases when the direction of cosmic muons is pointing to the interaction point. To
correctly and efficiently reconstruct cosmic muons and beam-halo muons, the cosmic muon
reconstruction algorithm assumes that muons are coming from outside, and is optimized by 
utilizing properties of cosmic muons and beam-halo muons as discussed below.

The new cosmic muon reconstruction software~\cite{algo} has been released and is available to the CMS
community as a part of the official CMS software releases. During the
MTCC the new cosmic muon reconstruction software was employed successfully to
reconstruct cosmic muons traversing a full slice of the CMS detector. Although the initial
motivation of the design was to reconstruct cosmic muons, the
reconstruction algorithm can also be applied to beam-halo muons.

\section{THE CMS MUON SYSTEM}
The CMS muon system~\cite{ptdr} is composed of 3 independent subsystems. 
In the barrel region ($|\eta|<0.8$), drift tube (DT) detectors are installed,
while cathode strip chambers (CSC) are
used in the endcap regions ($1.2<|\eta|<2.4$). 
In the intermediate (``overlap") region ($0.8<|\eta|<1.2$), chambers of
both detectors are crossed by a muon track from the interaction point.
Resistive plate chambers (RPC)
are installed in the $|\eta|<1.6$ region, covering both the barrel and the endcaps.
RPCs have limited spatial
resolution, but good time resolution, thus can provide excellent bunch crossing
identification.
The barrel muon system is arranged in 5 wheels along the $z$-axis, where each wheel is
divided into 12 sectors and 4 stations called (from innermost to
outermost) MB1, MB2, MB3, and MB4. Each station consists of 12 chambers, except for MB4, which has 14 chambers.
The endcap muon system is arranged in 4 stations at each end of the detector. They are
numbered from ME1 to ME4 in order of their absolute values of $z$-position. The
innermost CSC stations are composed of 3 concentric rings, while the other stations are
composed of 2 rings only. Each ring consists of 18 or 36 trapezoidal chambers.

\section{RESULTS}
\begin{figure*}[t]
\centering
\includegraphics[width=125mm]{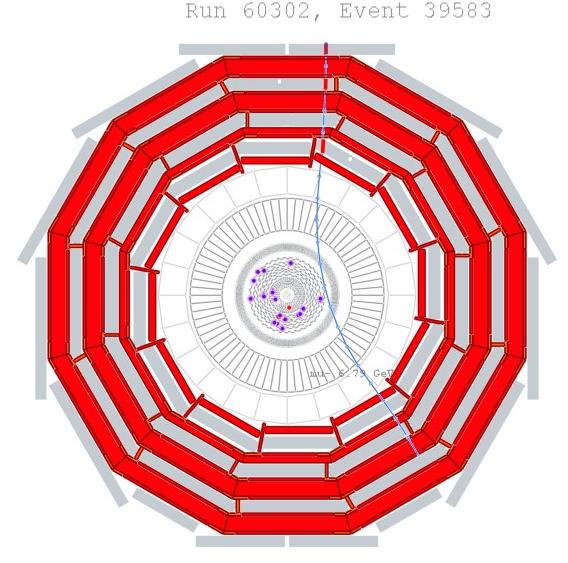}
\caption{Event display of a reconstructed cosmic muon.} \label{fig:cosmic}
\end{figure*}

The performance of the new cosmic muon reconstruction algorithm was studied using
simulated data from a dedicated Monte Carlo cosmic muon generator~\cite{philipp} as well as 
data taken during MTCC. 
Fig.~\ref{fig:cosmic} shows an event display of a reconstructed cosmic muon in a 3.8~T magnetic field. 

It is possible to observe muons that traverse the whole CMS detector,
as illustrated in Fig.~\ref{cospp}(b).
With the algorithm described here, 
all hits from both hemispheres of the detector can be used in 
a single trajectory, which in turn allows for a more precise momentum
measurement and provides an excellent tool for alignment.

The efficiency to reconstruct traversing muons is defined as
the number of events containing a track passing through 2 hemispheres 
divided by the number of events containing 2 separate standalone tracks in the 2
hemispheres. The measured efficiency is about 85\%.
Because more hits over a larger distance are included in 
traversing trajectories, 
the $p_T$ resolution of traversing tracks is better than for muons reconstructed only in one hemisphere.

\subsection{Global Cosmic Muon Reconstruction}
\label{sec:perfglobal}

The first reconstructed muon trajectory passing through the Tracker, DTs, and CSCs
 was reported shortly after the MTCC phase I. The efficiency of global cosmic muon reconstruction is defined as
the number of events with the global cosmic muon track built successfully over the
number of events with 1 standalone cosmic muon track and 1
tracker track. The measured efficiency by this
definition was to be about 46\%. 

\subsection{Strategy and Performance of Beam-Halo Muon Reconstruction}\label{beamhalo}

Beam-halo muons are machine-induced particles that travel along
the beam line from outside of the detector.
In CMS software, beam-halo muons can be reconstructed by the same
software package and configuration used for cosmic muons.
When the $|\eta|$ value of the momentum of
a trajectory seed in the endcap region exceeds a given threshold, it is identified as a
beam-halo muon. In this case, all barrel layers are skipped when asking
for compatible layers in the navigation step, 
because beam-halo muons will pass through the
entire sensitive zone of the barrel DT and RPC chambers, which creates a large
amount of charge to be deposited and decreases the chamber efficiency~\cite{ptdr}. 
Although not all beam-halo muons arrive at the second endcap,
the layers in the second endcap are still chosen as compatible layers.
The compatible layers are ordered as outside-in on one end and inside-out
on the other. The navigation direction is flipped from outside-in to inside-out
when the endcap region changes
during building trajectory. 
\begin{figure*}[t]
\centering
\includegraphics[width=135mm]{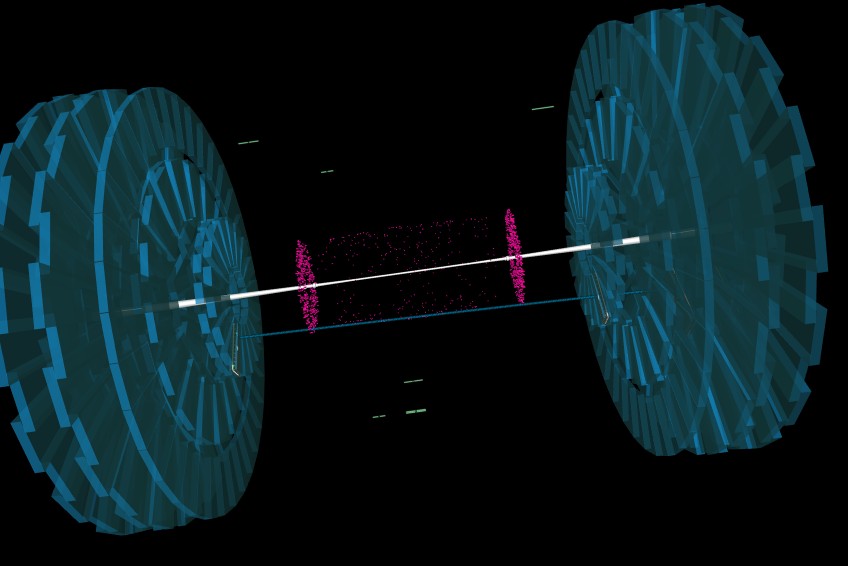}
\caption{Event display of a reconstructed beam-halo muon.} \label{fig:beamhalo}
\end{figure*}

Fig.~\ref{fig:beamhalo} shows a reconstructed beam-halo muon recorded during the LHC beam operation in September 2008.
The reconstruction efficiency, defined
as the number of events with track reconstructed
successfully divided by number of events with 2 or more track segments built
in the CSC system, is about 99\%.

\section{CONCLUSIONS} \label{conclusions}

We have described a new algorithm designed to reconstruct
cosmic muons and discussed the different reconstruction strategy compared to the
standard muon reconstruction algorithm. The new cosmic muon 
reconstruction algorithm works efficiently for both cosmic muons and 
beam-halo muons. A full detector simulation and reconstruction analysis was carried out to
validate the performance. In addition, data taken during the MTCC were compared to simulated cosmic data,
and good agreement between simulated and reconstructed results was observed. 
The presented cosmic muon reconstruction software provides a powerful tool to utilize cosmic
and beam-halo muons for synchronization and alignment during the commissioning of the
CMS detector and the initial data taking period at the LHC.

\end{document}